\documentclass[prb,twocolumn,showpacs,preprintnumbers,amsmath,amssymb]{revtex4}
\usepackage{epsf}
\usepackage{graphicx}
\usepackage{dcolumn}
\usepackage{bm}

\begin{document}

\title{Manifestation of spin-orbit interaction in tunneling between
2D electron layers}

\author{ I.~V.~Rozhansky}
\email{Second.Author@institution.edu}
\author{N.~S.~Averkiev}

\affiliation{A.F.Ioffe Physico-Technical Institute, Russian Academy
of Sciences, 194021 St.Petersburg, Russia}
\date{\today}

\begin{abstract}
{An influence of spin-orbit interaction on the tunneling between two
2D electron layers is considered. Particular attention is addressed
to the relation between the contribution of Rashba and Dresselhaus
types. It is shown that without scattering of the electrons, the
tunneling conductance can either exhibit resonances at certain
voltage values or be substantially suppressed over the whole voltage
range. The dependence of the conductance on voltage turns out to be
very sensitive to the relation between Rashba and Dresselhaus
contributions even in the absence of magnetic field. The elastic
scattering broadens the resonances in the first case and restores
the conductance to a larger magnitude in the latter one. These
effects open possibility to determine the parameters of spin-orbit
interaction and electrons scattering time in tunneling experiments
with no necessity of external magnetic field.}
\end{abstract}


\maketitle

\index{Rozhansky I. V.}
\index{Averkiev N. S.}
\section{Introduction}
Spin-orbit interaction (SOI) plays an important role in the widely
studied spin-related effects and spintronic devices. In the latter
it can be either directly utilized to create spatial separation of
the spin-polarized charge carries or indirectly influence the device
performance through spin-decoherence time. In 2D structures two
kinds of SOI are known to be of the most importance, namely Rashba
and Dresselhaus mechanisms. The first one characterized by parameter
$\alpha$ is due to the structure inversion asymmetry (SIA) while the
second one characterized by $\beta$ is due to the bulk inversion
asymmetry (BIA). Most brightly both of the contributions reveal
themselves when the values of $\alpha$ and $\beta$ are comparable.
In this case a number of interesting effects occur: the electron
energy spectrum becomes strongly anisotropic \cite{AnisotrSpectrum},
the electron spin relaxation rate becomes dependent on the spin
orientation in the plane of the quantum well
\cite{AverkievObserved}, a magnetic break-down should be observed in
the Shubnilov de Haas effect\cite{magn}. The energy spectra
splitting due to SOI can be observed in rather well-developed
experiments as that based on Shubnikov--de Haas effect. However,
these experiments can hardly tell about the partial contributions of
the two mechanisms leaving the determination of the relation between
$\alpha$ and $\beta$ to be a more challenging task. At the same
time, in some important cases spin relaxation time $\tau_s$ and spin
polarization strongly depend on the $\frac{\alpha}{\beta}$ ratio. In
this paper we consider the tunneling between 2D electron layers,
which turns out to be sensitive to the relation between Rashba and
Dresselhaus contributions. The specific feature of the tunneling in
the system under consideration is that the energy and in-plane
momentum conservation put tight restrictions on the tunneling.
Without SOI the tunneling conductance exhibits delta function-like
maximum at zero bias broadened by elastic scattering in the layers
\cite{MacDonald}, and fluctuations of the layers width
\cite{VaksoFluctuations}. Such a behavior was indeed observed in a
number of experiments \cite{Eisenstein,Turner,Dubrovski}. Spin-orbit
interaction splits the electron spectra into two subbands in each
layer. At that energy and momentum conservation can be fulfilled for
the tunneling between opposite subbands of the layers at a finite
voltage corresponding to the subbands splitting. However, if the
parameters of SOI are equal for left and right layers, the tunneling
remains prohibited due to orthogonality of the appropriate spinor
eigenstates. In \cite{Raichev} it was pointed out that this
restriction can also be eliminated if Rashba parameters are
different for the two layers. A structure design was proposed
\cite{Raikh} where exactly opposite values of the Rashba parameters
result from the built-in electric field in the left layer being
opposite to that in the right layer. Because the SOI of Rashba type
is proportional to the electric field, this would result in
$\alpha^R=-\alpha^L$, where $\alpha^L$ and $\alpha^R$ are the Rashba
parameters for the left and right layers respectively. In this case
the
 peak of the conductance should occur at the voltage $U_0$ corresponding
to the energy of SOI: $eU_0=\pm2\alpha k_F$, where $k_F$ is Fermi
wavevector. In this paper we consider arbitrary Rashba and
Dresselhaus contributions and show how qualitatively different
situations can be realized depending on their partial impact. In
some cases the structure of the electrons eigenstates suppresses
tunneling at ever voltage. At that the scattering is important as it
restores the features of voltage-current characteristic containing
information about SOI parameters. Finally the parameters $\alpha$
and $\beta$ can be obtained in the tunneling experiment which unlike
other spin-related experiments requires neither magnetic field nor
polarized light.
\section{Calculations}
We consider two 2D electron layers separated by potential barrier at
zero temperature (see Fig.\ref{fig:layers}). We shall consider only
one level of size quantization and not too narrow barrier so that
the electrons wavefunctions in the left and right layers overlap
weakly.
The system can be described by the phenomenological tunneling Hamiltonian \cite{MacDonald,MacDonald2,VaksoFluctuations}:
\begin{figure}[h]
\leavevmode
 \centering\epsfxsize=180pt \epsfbox[30 530 500 760]{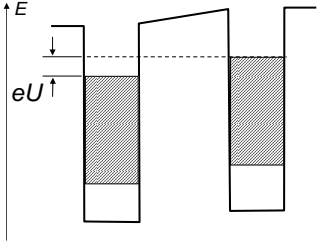}
\caption{\label{fig:layers} Energy diagram of two 2D electron
layers.}
\end{figure}
\begin{equation}
\label{HT0} H=H_{0}^L+H_{0}^R+H_T,
\end{equation}
where $H_{0}^L,H_{0}^R$ are the partial Hamiltonians for the left
and right layers respectively, $H_T$ is the tunneling term. With
account for the elastic scattering and SOI in the layers the partial
Hamiltonians and the tunneling term have the the following form in
representation of secondary quantization:
\begin{equation}
\label{eqH}
\begin{array}{l}
H_{0}^l = \sum\limits_{k,\sigma} {\varepsilon^l_{k} c^{l+}_{k\sigma}
 c^l_{k\sigma } } +  \sum\limits_{k,k',\sigma}  {V^l_{kk'} c^{l+}_{k\sigma}c^l_{k'\sigma }} + H^l_{SO}  \\
H_T  = \sum\limits_{k,k',\sigma,\sigma'} {T_{kk'\sigma\sigma'}\left( {c^{L+}_{k\sigma} c^{R}_{k'\sigma'}  + c^{R+}_{k'\sigma'}  c^L_{k\sigma} } \right)}, \\
 \end{array}
\end{equation}
Here index $l$ is used for the layer designation and can take the
values $l=R$ for the right layer, $l=L$ for the left layer. By $k$
here and further throughout the paper we denote the wavevector
aligned parallel to the layers planes, $\sigma$ denotes spin
polarization and can take the values $\sigma=\pm 1/2$.
$\varepsilon_k^l$ is the energy of an electron in the layer $l$
having in-plane wavevector $k$. It can be expressed as:
\begin{equation}
\label{spectrum}
 \varepsilon _k^l  = \varepsilon+\varepsilon_0^l+\Delta^l,
 \end{equation}
where $\varepsilon=\frac{\hbar^2k^2}{2m}$, $m$ being electron's
effective mass, $\varepsilon_0^l$ and $\Delta^l$ are the size
quantization energy and the energy shift due to external voltage for
the layer $l$ . We shall also use the value $\Delta^{ll'}$ defined
as
$\Delta^{ll'}=(\Delta^l-\Delta^{l'})+(\varepsilon_0^l-\varepsilon_0^{l'})$.
Similar
 notation will be used for spin polarization denoted by indices $\sigma$, $\sigma'$.
 The second term in  the Hamiltonian (\ref{eqH}) $V_{kk'}^l$ is the matrix element of the scattering operator.
  We consider only elastic scattering. The tunneling
term $H_T$ in (\ref{eqH}) is described by the tunneling constant
$T_{kk'\sigma\sigma'}$, which
 has the meaning of size quantization levels splitting due to
the wavefunctions overlap. By lowercase $t$ we shall denote the
overlap integral itself. Our consideration is valid only for the
case of weak overlapping, i.e. $t\ll1$. Parametrically $T\sim
t\varepsilon_F$, where $\varepsilon_F$ is the electrons Fermi
energy.  The term $H^{l}_{SO}$ describes the spin-orbit part of the
Hamiltonian:
\begin{equation}
\label{eqSOH}
 \hat{H}^l_{SO}=\alpha^l \left( \bm{\sigma}  \times \bm{k}
\right)_z + \beta^{l} \left( {\sigma _x k_x  - \sigma _y k_y }
\right),
\end{equation}
where $\sigma_i$ are the Pauli matrices, $\alpha^l,\beta^l$ are
respectively the parameters of Rashba and Dresselhaus interactions
for the layer $l$. In the secondary quantization representation:
\begin{eqnarray}
 \hat {H}_{SO}^l =\alpha^l \sum\limits_k {\left( {k_y
-ik_x } \right)c_{k\sigma }^{l+} c_{k\sigma '}^l +} \left( {k_y
+ik_x }
\right)c_{k\sigma '}^{l+} c_{k,\sigma }^l \nonumber \\
 +\beta^l \sum\limits_k
{\left( {k_x -ik_y } \right)c_{k\sigma }^{l+} c_{k\sigma '}^l +}
\left( {k_x +ik_y } \right)c_{k\sigma '}^{l+} c_{k\sigma }^l
\label{eqSOHc}
\end{eqnarray}
The operator of the tunneling current can be expressed as
\cite{MacDonald}:
\begin{equation}
\label{current0}
 \hat{I} = \frac{{ie}}{\hbar
}\sum\limits_{k,k',\sigma,\sigma'} T_{kk'\sigma\sigma'}
\left(\hat\rho_{kk'\sigma\sigma'}^{RL}-\hat\rho_{k'k\sigma'\sigma}^{LR}
\right),
\end{equation}
where
$\hat\rho_{kk'\sigma\sigma'}^{ll'}=c_{k,\sigma}^{l+}c_{k',\sigma'}^{l'}$
 We shall assume the case of in-plane momentum and the spin
projection being conserved in the tunneling event so the tunneling
constant $T_{kk'\sigma\sigma'}$ has the form
$T_{kk'\sigma\sigma'}=T\delta_{kk'}\delta_{\sigma\sigma'}$, where
$\delta$ is the Cronecker symbol. The tunneling current is then
given by
 \begin{equation}
 \label{current}
 I = \frac{ie}{\hbar}
T \int dk\: \mathrm{Tr} \left( \left<\hat\rho^{RL}_{k\sigma}\right>
-\left<\hat\rho^{LR}_{k\sigma}\right>\right),
\end{equation}
where $<>$ denotes the expectation value in quantum-mechanical
sense. For further calculations it is convenient to introduce vector
operator
$\bm{\hat{S}}^{ll'}_{kk'}=\left\{\hat{S_0},\bm{\hat{s}}\right\}=\left\{\mathrm{Tr}\left(\hat\rho^{ll'}_{kk'\sigma\sigma'}\right),\mathrm{Tr}\left({\bm
\sigma}\hat\rho^{ll'}_{kk'\sigma\sigma'}\right) \right\}$. This
vector fully determines the current because the latter can be
expressed through the difference
$\hat{S}^{RL}_{0k}-\hat{S}^{LR}_{0k}$. The time evolution of
$\bm{\hat{S}}^{ll'}_{kk'}$ is governed by:
\begin{equation}
\label{drodt}
\frac{d\bm{\hat{S}}_{kk'}^{ll'}}{dt}=\frac{i}{\hbar}[H,\bm{\hat{S}}_{kk'}^{ll'}]
\end{equation}
In the standard way of reasoning \cite{Luttinger} we assume
adiabatic onset of the interaction with characteristic time
$w^{-1}$. We will set $w=0$ in the final expression. With this
(\ref{drodt}) turns into:
\begin{equation}
\label{drodt0}
(\bm{\hat{S}}_{kk'}^{ll'}-\bm{\hat{S}}_{kk'}^{(0)ll'})w=\frac{i}{\hbar}[H,\bm{\hat{S}}_{kk'}^{ll'}]
\end{equation}
Here $\bm{\hat{S}}_{kk'}^{(0)ll'}$ represents the stationary
solution of (\ref{drodt}) without interaction. By interaction here
we mean the tunneling and the elastic scattering by impurities but
not the external voltage. The role of the latter is merely shifting
the layers by $eU$ on the energy scale. From such defined
interaction it immediately follows that the only non-zero elements
of $\bm{\hat{S}}_{kk'}^{(0)ll'}$ are that with $l=l'$ and $k=k'$. In
further abbreviations we will avoid duplication of the indices i.e.
write single $l$ instead of $ll$ and $k$ instead of $kk$:
\begin{equation}
\label{Sdiag}
\bm{\hat{S}}_{kk'}^{(0)ll'}=\bm{\hat{S}}_{k}^{(0)l}\delta_{kk'}\delta_{ll'}
\end{equation}
 With use of fermion commutation rules
 \begin{eqnarray*}
 \left\{ {c_i c_k } \right\} = \left\{ {c_i^ +  c_k^ +  } \right\} = 0 \\
 \left\{ {c_i c_k^ +  } \right\} = \delta _{ik}
 \end{eqnarray*}
the calculations performed in a way similar to \cite{Luttinger}
bring us to the following system of equations
 with respect to
$\bm{\hat{S}}_{k}^{ll'}$:
\begin{eqnarray}
 0= \left( {\Delta^{ll'}+i\hbar w } \right){\bf{\hat
S}}_k^{ll'} + T\left( {{\bf{\hat S}}_k^{l'}  - {\bf{\hat S}}_k^l }
\right)+{\bf{M(}}k{\bf{)\hat S}}_k^{ll'} \nonumber \\
- \sum\limits_{k'} {\left( {\frac{{A_{kk'} {\bf{\hat S}}_k^{ll'}  -
B_{kk'} {\bf{\hat S}}_{k'}^{ll'} }}{{ {\varepsilon'  - \varepsilon
-\Delta^{ll'} }  + i\hbar w}} + \frac{{B_{kk'} {\bf{\hat S}}_k^{ll'}
- A_{kk'} {\bf{\hat S}}_{k'}^{ll'} }}{{ {\varepsilon   -
\varepsilon' -\Delta^{ll'} }  + i\hbar w}}} \right)}
 \label{system1}
 \end{eqnarray}
\begin{eqnarray}
i\hbar w\left( {{\bf{\hat S}}_k^{\left( 0 \right)l} - {\bf{\hat
S}}_k^l } \right) = T\left( {{\bf{\hat S}}_k^{l'l}  - {\bf{\hat
S}}_k^{ll'} } \right) + {\bf{M}}(k){\bf{\hat S}}_k^l \nonumber \\ +
\sum\limits_{k'} { {\frac{{2i\hbar wA_{kk'} \left( {{\bf{\hat
S}}_k^l  - {\bf{\hat S}}_{k'}^{l'} } \right)}}{{\left( {\varepsilon'
- \varepsilon } \right)^2  + \left( {\hbar w} \right)^2 }}} },
\label{system2}
\end{eqnarray}
where $\bm{M}$ is a known matrix, depending on $k$ and parameters of
spin-orbit interaction in the layers. Here we also introduced the
quadratic forms of the impurities potential matrix elements:
\begin{eqnarray}
A_{kk'}  \equiv \left| {V_{k'k}^{l} } \right|^2 \nonumber \\
B_{kk'} \equiv V_{k'k}^{l} V_{kk'}^{l'}
\label{correlators}
\end{eqnarray}
As (\ref{system1}) and (\ref{system2}) comprise a system of linear
integral equations these quantities enter the expression
(\ref{current}) for the current linearly and can be themselves
averaged over spatial distribution of the impurities. In order to
perform this averaging we assume the short range potential of
impurities:
\begin{equation}
\label{ImpuritiesPotential} V\left( r \right) = \sum\limits_a
{V_0^{} \delta \left( {r - r_a } \right)}
\end{equation}
The averaging immediately shows that the correlators
$\left<A_{kk'}\right>\equiv A$ and $\left<B_{kk'}\right>\equiv B$
have different parametrical dependence on the tunneling transparency
$t$, namely
\begin{equation}
\label{T2}
 \frac{B}{A}\sim t^{2}\sim T^2
 \end{equation}
We emphasize that this result holds for non-correlated distribution
of the impurities as well as for their strongly correlated
arrangement such as a thin layer of impurities placed in the middle
of the barrier. The corresponding expressions for these two cases
are given below. Index 'rand' stands for uniform impurities
distribution and 'cor' for their correlated arrangement in the
middle of the barrier $(z=0)$:
\begin{eqnarray}
 {B^{rand} }  = \frac{{V_0^2 n}}{W}\int {dz}
f_l ^2 (z)f_{l'} ^2 (z)\sim\frac{{V_0^2
n}}{W}\frac{{t^2 }}{d} \nonumber \\
  {A^{rand} }
  = \frac{{V_0^2 n}}{W}\int {dz} f_l^4\left(z\right)
\sim\frac{{V_0^2 n}}{W}\frac{1}{d}
\nonumber \\
{B^{cor} }   = \frac{{V_0^2 n_s }}{W}f_l ^2 (0)f_{l'} ^2
(0)\sim\frac{{V_0^2 n_s}}{W}\frac{{t^2 }}{d}
\nonumber \\
 {A^{cor} }   = \frac{{V_0^2 n_s
}}{W}f_l ^4 \left( 0 \right)\sim\frac{{V_0^2 n_s}}{W}\frac{1}{d},
\label{correlators1}
\end{eqnarray}
where $n$ and $n_s$ are bulk and surface concentrations of the
impurities, $W$ is the lateral area of the layers, $d$ is the width
of the barrier and $f(z)$ is the eigenfunction corresponding to the
size quantization level, $z$ is coordinate in the direction normal
to the layers planes, $z=0$ corresponding to the middle of the
barrier\cite{Raikh}.
 Unlike \cite{Raikh} and according to (\ref{T2}) we
conclude that the correlator $\left<B_{kk'}\right>$ has to be
neglected as soon as we shall be interested in calculating the
current within the order of $T^2$. In the hereused method of
calculation this result appears quite naturally, however, it can be
similarly traced in the technique used in \cite{Raikh} (see
Appendix). For the same reason the tunneling term should be dropped
from (\ref{system2}) as it would give second order in $T$ if
(\ref{system2}) substituted into (\ref{system1}). According to
(\ref{correlators}) $A$ can be expressed in terms of electrons
scattering time:
\begin{equation}
\label{tau} \frac{1}{\tau } = \frac{{2\pi }}{\hbar }\nu\left\langle
{\left| {V_{kk'} } \right|^2 } \right\rangle   = \frac{{2\pi
}}{\hbar }\nu A ,
\end{equation}
where $\nu$ is the 2D density of states $\nu=\frac{m}{2\pi\hbar^2}$.
By means of Fourier transformation on energy variable the system
(\ref{system1}),(\ref{system2}) can be reduced to the system of
linear algebraic equations. Finally ${{\bf{\hat S}}_k^{ll'} }$ can
be expressed as a function of ${{\bf{\hat S}}_k^{\left( 0 \right)l}
}$. Consequently the current (\ref{current}) becomes a function of
$\left<\hat{\rho}_{k\sigma}^{(0)R}\right>$,
$\left<\hat{\rho}_{k\sigma}^{(0)L}\right>$. For the considered case
of zero temperature:
    \[
\left<\rho _{k\sigma}^{(0)l}\right>  = \frac{1}{2W}  \theta \left(
{\varepsilon _F^l  + \Delta ^l - \varepsilon - \varepsilon _\sigma }
\right),
\]
where
\[
\varepsilon _\sigma   =  \pm \left| {\alpha ^l \left( {k_x  - ik_y }
\right) - \beta ^l \left( {ik_x  - k_y } \right)} \right|,
\]. Without loss of generality we shall consider the
case of identical layers and external voltage applied as shown in
Fig.\ref{fig:layers}:
\begin{eqnarray*}
\varepsilon_0^R=\varepsilon_0^L\\
\Delta^L=-\frac{eU}{2}, \Delta^R=+\frac{eU}{2}\\
 \Delta^{RL}=-\Delta^{LR}=eU
\end{eqnarray*}
 The calculations can be simplified with account for
two small parameters:
\begin{eqnarray}
 \xi=\frac{\hbar}{\varepsilon_F\tau}\ll1  \nonumber \\
\eta=\frac{eU}{\varepsilon_F}\ll1 \label{deltaef}
\end{eqnarray}
With (\ref{eftau}) calculation yields the following expression for
the current:
\begin{equation}
\label{currentfinal0}  I = \frac{{ie}}{{2\pi \hbar }}T^2 \nu
\int\limits_0^\infty {\int\limits_0^{2\pi } {\left( {\zeta ^L  +
\zeta ^R } \right)\mathrm{Tr}\left( {\rho _\sigma ^{\left( 0
\right)R}  - \rho _\sigma ^{\left( 0 \right)L} } \right)d\varepsilon
d\varphi } },
\end{equation}
where
\[
\label{constants} \zeta ^l = \frac{{C^l \left[ {\left( {C^l }
\right)^2 - 2bk^2 \sin2\varphi  - gk^2 } \right]}}{{\mathop {\left(
{f + 2d\sin2\varphi } \right)}\nolimits^2 k^4  - 2\left( {C^l }
\right)^2 \left( {c + 2a\sin2\varphi } \right)k^2  + \left( {C^l }
\right)^4 }}, \]
 \[ C^l\left(U\right)  = \Delta ^l  + i\frac{\hbar
}{\tau },
\]
\begin{eqnarray}
 a  = \alpha ^L \beta ^L  + \alpha ^R \beta ^R \nonumber  \\
 b  = \left( {\beta ^L  + \beta ^R } \right)\left( {\alpha ^L  + \alpha ^R } \right)\nonumber \\
 c  = \left( {\beta ^L } \right)^2  + \left( {\beta ^R } \right)^2  + \left( {\alpha ^L } \right)^2  + \left( {\alpha ^R } \right)^2 \nonumber \\
 d  = \alpha ^L \beta ^L  - \alpha ^R \beta ^R \nonumber \\
 f  = \left( {\beta ^L } \right)^2  - \left( {\beta ^R } \right)^2  + \left( {\alpha ^L } \right)^2  - \left( {\alpha ^R } \right)^2 \nonumber \\
 g  = \mathop {\left( {\beta ^L  + \beta ^R } \right)}\nolimits^2  + \mathop {\left( {\alpha ^L  + \alpha ^R } \right)}\nolimits^2  \nonumber \\
\label{constants}
 \end{eqnarray}
Parameters $a$-$g$ are various combinations of the Rashba and
Dresselhaus parameters of SOI in the layers. Both types of SOI are
known to be small in real structures so that:
\begin{equation}
\alpha k_F\ll\varepsilon_F, \; \beta k_F\ll\varepsilon_F
\end{equation}
This additional assumption together with (\ref{deltaef}) reduces
(\ref{currentfinal0}) to
\begin{equation}
\label{currentfinal}  I =  \frac{{ie^2 }}{{2\pi \hbar }}T^2 \nu
WU\int\limits_0^{2\pi } {\left[ {\zeta ^L \left( {\varepsilon_F }
\right) + \zeta ^R \left( {\varepsilon_F } \right)} \right]d\varphi
}
\end{equation}
The integral over $\varphi$ in (\ref{currentfinal}) can be
calculated analytically by means of complex variable integration.
However, the final result for arbitrary $\alpha^l,\beta^l$  is not
given here for it is rather cumbersome. In the next section some
particular cases are discussed.
\section{Results and Discussion}
The obtained general expression (\ref{currentfinal}) can be
simplified for a few particular important relations between Rashba
and Dresselhaus contributions. These calculations reveal
qualitatively different dependencies of the d.c. tunneling current
on the applied voltage.
\begin{figure}[h]
 \leavevmode
 \centering\epsfxsize=210pt \epsfbox[130 350 700 800]{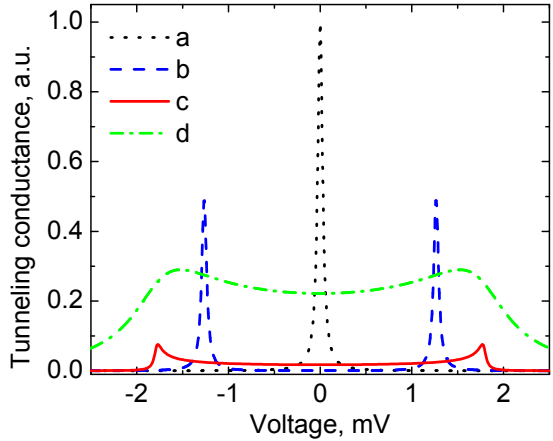}
\caption{\label{fig:tunnelingmain}Tunneling conductance, a:
$\varepsilon_F=10$ meV, $\alpha=\beta=0$, $\tau=2*10^{-11}$ s; b:
same as a, but $\alpha k_F=0.6$ meV; c: same as b, but
$\beta=\alpha$; d: same as c, but $\tau=2*10^{-12}$ s.}
\end{figure}
The results of the calculations shown below were obtained using the
following parameters: Fermi energy $\varepsilon_F=10$ meV,
spin-orbit splitting was taken to resemble $GaAs$ structures:
$\alpha k_F=0.6$ meV.
\subsection{No Spin-Orbit Interaction}
 In the absence of SOI ($\alpha^R=\alpha^L=0$, $\beta^R=\beta^L=0$) the
energy spectrum for each of the layers forms a paraboloid:
\begin{equation}
E^l(k)=\varepsilon_0+\frac{\hbar^2k^2}{2m}\pm \frac{eU}{2}.
\end{equation}
  According to our assumptions (\ref{current0}),(\ref{current}), the tunneling takes place at:
\begin{eqnarray}
   E^R=E^L\nonumber \\
   k^R=k^L
   \label{conservation}
\end{eqnarray}
Both conditions are satisfied
  only at $U=0$ so that a nonzero external voltage does not produce any current
  despite it produces empty states in one layers aligned to the filled states in the other layer
  (Fig.\ref{fig:layers}). The momentum conservation restriction in (\ref{conservation}) is weakened if the electrons scatter at the impurities.
   Accordingly, one should expect a nonzero tunneling current
  within a finite voltage range in vicinity of zero.
 For the considered case the general formula (\ref{currentfinal}) is simplified radically as all the parameters (\ref{constants})
 have zero values. Finally we get the well-known
 result\cite{MacDonald}:
\begin{equation}
\label{currentMacDonald}
 I = 2e^2 T^2 \nu
WU\frac{{\frac{1}{\tau }}}{{\left( {eU} \right)^2 + \left(
{\frac{\hbar }{\tau }} \right)^2 }}. \end{equation}
 The conductance defined as $G(U)=I/U$ has  Lorentz-shaped peak at $U=0$
 turning into delta function at $\tau\rightarrow\infty$.
 This case is shown in (Fig.\ref{fig:tunnelingmain},a).
  All the curves in Fig.\ref{fig:tunnelingmain} show the results of the
calculations for very weak scattering. The corresponding scattering
time is taken $\tau=2*10^{-11}s$.
\subsection{Spin-Orbit Interaction of Rashba type}
\begin{figure}[h]
 \leavevmode
 \centering\epsfxsize=220pt \epsfbox[130 500 700 800]{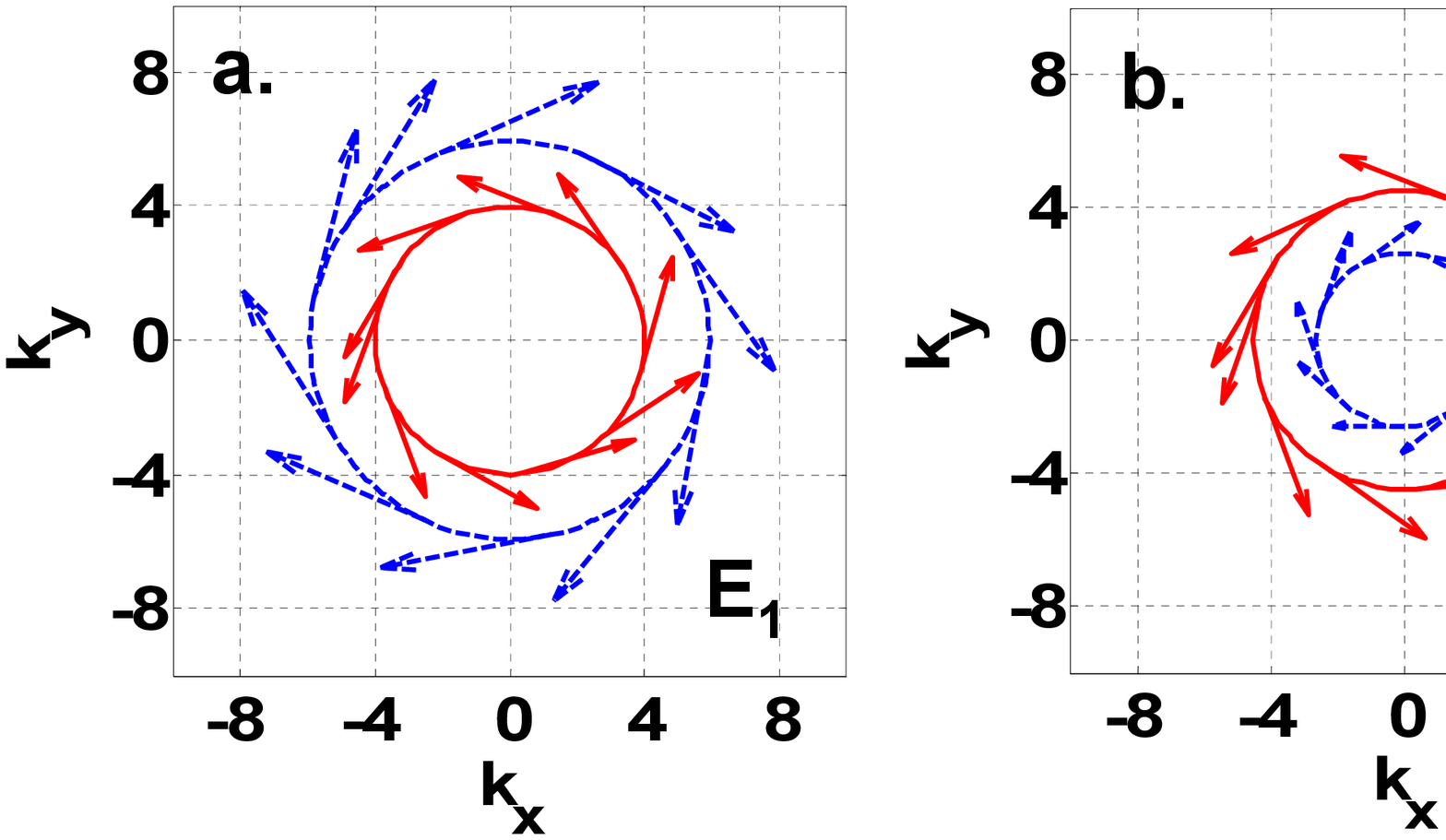}
 \caption{\label{fig:spectraRashba}Cross-section of electron energy spectra in the left(a) and right (b) layer for
 the case
$\alpha^{L}=-\alpha^{R}, \beta^{L}=\beta^{R}=0$.}
\end{figure}

 The spin-orbit interaction gives
qualitatively new option for the d.c. conductance to be finite at
non-zero voltage. SOI splits the spectra into two subbands. Now an
electron from the first subband of the left layer can tunnel to a
state in a second subband of the right layer. Let us consider a
particular case when only Rashba type of SOI interaction exists in
the system, its magnitude being the same in both layers, i.e.
$|\alpha^R|=|\alpha^L|\equiv \alpha$, $\beta^R=\beta^L=0$. In this
case the spectra splits into two paraboloid-like subbands "inserted"
into each other. Fig.\ref{fig:spectraRashba} shows their
cross-sections for both layers,
 arrows show spin orientation. By applying a certain external
voltage $U_0=\frac{2\alpha k_F}{e}$,
$k_F=\frac{\sqrt{2m\varepsilon_F}}{\hbar}$ the layers can be shifted
on the energy scale in such a way that the cross-section of the
"outer" subband of the right layer coincides with the "inner"
subband of the left layer (see solid circles in
Fig.\ref{fig:spectraRashba}). At that both conditions
(\ref{conservation}) are satisfied. However, if the spin is taken
into account, the interlayer transition can still remain forbidden.
It happens if the appropriate spinor eigenstates involved in the
transition are orthogonal. This very case occurs if
$\alpha^R=\alpha^L$, consequently the conductance behavior remains
the same as that without SOI. Contrary, if the Rashba terms are of
the opposite signs, i.e. $\alpha^R=-\alpha^L$ the spin orientations
in the "outer" subband of the right layer and the "inner" subband of
the left layer are the same and the tunneling is allowed at a finite
voltage but forbidden at $U=0$ . This situation, pointed out in
\cite{Raichev,Raikh} should reveal itself in sharp maxima of the
conductance at $U=\pm U_0$ as shown in
Fig.\ref{fig:tunnelingmain},b. From this dependence the value of
$\alpha$ can be immediately extracted from the position of the peak.
Evaluating (\ref{constants}) for this case and further the
expression (\ref{currentfinal}) we obtain the following result for
the current:
\begin{equation}
\label{currentRaikh}  I = \frac{{2e^2T^2 W\nu U\frac{\hbar }{\tau
}\left[ {\delta^2 + e^2 U^2  + \left( {\frac{\hbar }{\tau }}
\right)^2 } \right]}}{{\left[ {\left( {eU - \delta } \right)^2 +
\left( {\frac{\hbar }{\tau }} \right)^2 } \right]\left[ {\left( {eU
+ \delta } \right)^2  + \left( {\frac{\hbar }{\tau }} \right)^2 }
\right]}},
\end{equation}
where $\delta=2\alpha k_F$. The result is in agreement with that
derived in\cite{Raikh}, taken for uncorrelated spatial arrangement
of the impurities. As we have already noted we do not take into
account interlayer correlator $\left<B_{kk'}\right>$
($\ref{correlators}$) because parametrically it has higher order of
tunneling overlap integral $t$ than the intralayer correlator
$\left<A_{kk'}\right>$. Therefore the result (\ref{currentRaikh}) is
valid for arbitrary degree of correlation in spatial distribution of
the impurities in the system.
It is worth noting that the opposite case when only Dresselhaus type
of SOI exists in the system leads to the same results. However, it
is rather non-practical to study the case of the different
Dresselhaus parameters in the layers because this type of SOI
originates from the crystallographic asymmetry and therefore cannot
be varied if the structure composition is fixed. For this case to be
realized one needs no make the two layers of different materials.
\subsection{Both Rashba and Dresselhaus contributions}
\begin{figure}[h]
 \leavevmode
 \centering\epsfxsize=190pt \epsfbox[130 350 700 800]{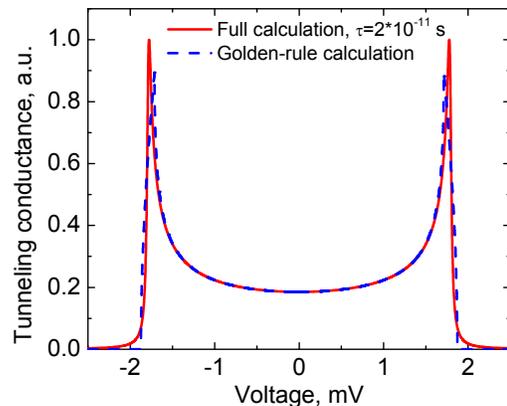}
\caption{\label{fig:goldenRule}Tunneling conductance calculated for
the case $\alpha^R=-\alpha^L=\beta$ and very weak scattering
compared to the precise result obtained through Fermi's golden rule
calculation.}
\end{figure}

\begin{figure}[h]
  \hfill
  \begin{minipage}[t]{0.5\textwidth}
    \begin{center}
      \centering\epsfxsize=170pt \epsfbox[70 650 266 801]{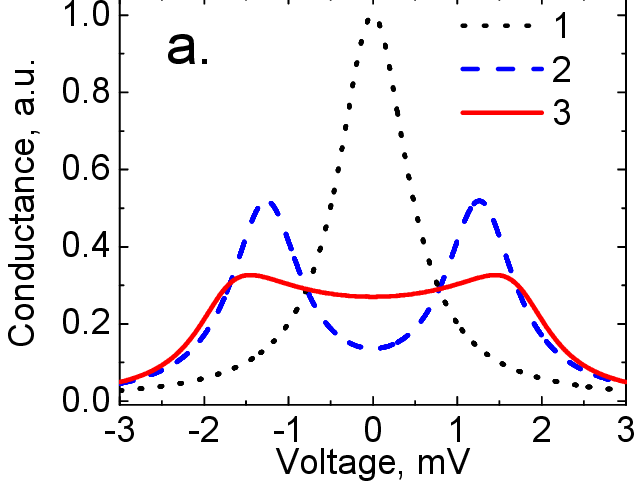}
    \nonumber
    \end{center}
  \end{minipage}
  \hfill
  \begin{minipage}[t]{0.5\textwidth}
    \begin{center}
      \epsfxsize=170pt \epsfbox[70 650 266 801]{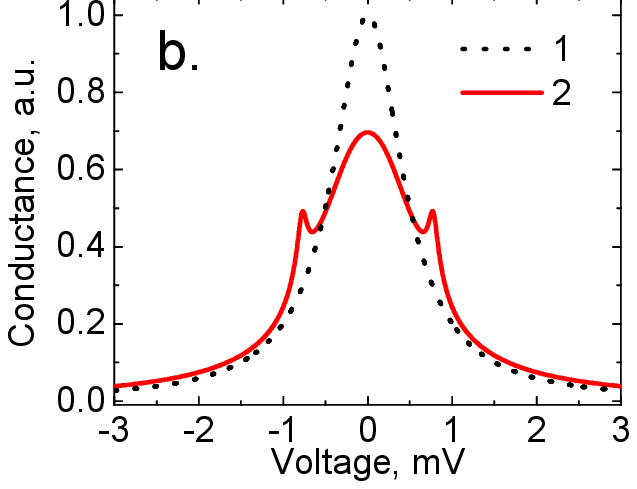}
      \nonumber
    \end{center}
  \end{minipage}
  \hfill
  \begin{minipage}[t]{0.5\textwidth}
    \begin{center}
      \epsfxsize=170pt \epsfbox[70 650 266 801]{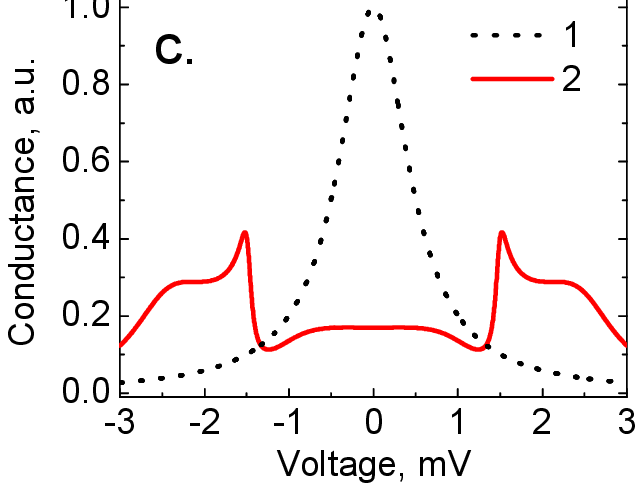}
    \end{center}
  \end{minipage}
  \caption{\label{fig:variousRD}Tunneling conductance calculated for various parameters of
SOI}
\end{figure}

The presence of Dresselhaus term in addition to the Rashba
interaction can further modify the tunneling conductance in a
non-trivial way. A special case occurs if the magnitude of the
Dresselhaus term is comparable to that of the Rashba term. We shall
always assume the Dresselhaus contribution being the same in both
layers: $\beta^{L}=\beta^{R}\equiv\beta$. Let us add the Dresselhaus
contribution to the previously discussed case so that
$\alpha^{L}=-\alpha^{R}\equiv\alpha,\;\alpha=\beta$. The
corresponding energy spectra and spin orientations are shown in
Fig.\ref{fig:spectraRD}. Note that while the spin orientations in
the initial and final states are orthogonal for any transition
between the layers, the spinor eigenstates are not, so that the
transitions are allowed whenever the momentum and energy
conservation requirement (\ref{conservation}) is fulfilled. It can
be also clearly seen from Fig.\ref{fig:spectraRD} that the condition
(\ref{conservation}), meaning overlap of the cross-sections a. and
b. occurs only at few points. This is unlike the previously
discussed case where the overlapping occurred within the whole
circular cross-section shown by solid lines in
Fig.\ref{fig:spectraRashba}. One should naturally expect the
conductance for the case presently discussed to be substantially
lower. Using (\ref{currentfinal}) we arrive at a rather cumbersome
expression for the current:
\begin{figure}[h]
 \leavevmode
 \centering\epsfxsize=220pt \epsfbox[130 500 700 810]{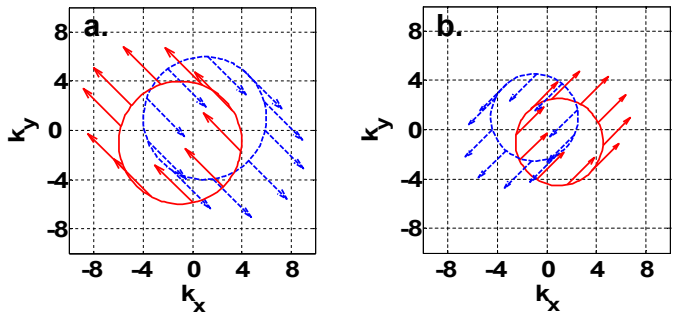}
\caption{\label{fig:spectraRD}Cross-section of electron energy
spectra in the left(a) and right (b) layer for
 the case
$\alpha^{R}=-\alpha^L=\beta$.}
\end{figure}
\begin{eqnarray}
   I = eT^2 W\nu U\left[ {\frac{{G_ - \left(
{G_ - ^2  - \delta ^2 } \right)}}{{\sqrt {F_ -  \left( {\delta ^4  +
F_ -  } \right)} }} - \frac{{G_ +  \left( {G_ + ^2  - \delta ^2 }
\right)}}{{\sqrt {F_ +  \left( {\delta ^4  + F_ +  } \right)} }}}
\right], \label{CurrentSpecial}
\end{eqnarray}
where \begin{eqnarray*}
G_ \pm   = eU \pm i\frac{\hbar }{\tau } \\
F_ \pm   = G_ \pm ^2 \left( {G_ \pm ^2  - 2\delta^2 } \right).
\end{eqnarray*}
Alternatively, for the case of no interaction with impurities a
precise formula for the transition rate between the layers can be
obtained by means of Fermi's golden rule. We obtained the following
expression for the current:
\begin{equation}
\label{CurrentPrecise} I = \frac{{2\pi eT^2 W}}{{\hbar \alpha ^2
}}\left( {\sqrt {K + \frac{{8m\alpha ^2 eU}}{{\hbar ^2 }}}  - \sqrt
{K - \frac{{8m\alpha ^2 eU}}{{\hbar ^2 }}} } \right),
\end{equation} where
\[
K = 2\delta^2  - e^2 U^2  + \frac{{16m^2 \alpha ^4 }}{{\hbar ^4 }}
\]
Comparing the results obtained from (\ref{CurrentSpecial}) and
(\ref{CurrentPrecise}) is an additional test for the correctness of
(\ref{CurrentSpecial}). Both dependencies are presented in
Fig.\ref{fig:goldenRule} and show a good match. The same dependence
of conductance on voltage is shown in Fig.\ref{fig:tunnelingmain},c.
As can be clearly seen in the figure the conductance is indeed
substantially suppressed in the whole voltage range. This is
qualitatively different from all previously mentioned cases.
Furthermore, the role of the scattering at impurities appears to be
different as well. For the considered above cases characterized by
resonance behavior of the conductance, the scattering broadens the
resonances into Lorentz-shape peaks with the characteristic width
$\delta=\hbar/(e\tau)$. Contrary, for the last case the weakening of
momentum conservation, caused by the scattering, increases the
conductivity and restores the manifestation of SOI in its dependence
on voltage. Fig.\ref{fig:tunnelingmain},d shows this dependence for
a shorter scattering time $\tau=2*10^{-12}$. The reason for that is
the weakening of the momentum conservation requirement due to the
elastic scattering. One should now consider the overlap of the
spectra cross-sections the circles in Fig.\ref{fig:spectraRD} having
a certain thickness proportional to $\tau^{-1}$. This increases the
number of points at which the overlap occurs and, consequently, the
value of the tunneling current. As the calculations show, for
arbitrary $\alpha$ and $\beta$ the dependence of conductance on
voltage can exhibit various complicated shapes with a number of
maxima, being very sensitive to the relation between the two
contributions. The origin of such a sensitivity is the interference
of the angular dependencies of the spinor eigenstates in the layers.
A few examples of such interference are shown in
Fig.\ref{fig:variousRD}, a--c. All the dependencies shown were
calculated for the scattering time $\tau=2*10^{-12}$ s.
Fig.\ref{fig:variousRD},a summarizes the results for all previously
discussed cases of SOI parameters, i.e. no SOI (curve 1), the case
$\alpha_R=-\alpha_L, \beta=0$ (curve 2) and
$\alpha_R=-\alpha_L=\beta$ (curve 3). Following the magnitude of
$\tau$ all the reasonances are broadenered compared to that shown in
Fig.\ref{fig:tunnelingmain}. Fig.\ref{fig:variousRD},b (curve 2)
demonstrates the conductiance calculated for the case
$\alpha_L=-\frac{1}{2}\alpha_R=\beta$, Fig.\ref{fig:variousRD},c
(curve 2) -- for the case $\alpha_L=\frac{1}{2}\alpha_R=\beta$. The
curve 1 corresponding to the case of no SOI is also shown in all the
figures for reference. Despite of a significant scattering parameter
all the patterns shown in Fig.\ref{fig:variousRD} remain very
distinctive. That means that in principle the relation between the
Rashba and Dresselhaus contributions to SOI can be extracted merely
from the I-V curve measured in a proper tunneling experiment.
\section{Summary}
As we have shown, in the system of two 2D electron layers separated
by a potential barrier SOI can reveal itself in the tunneling
current. The difference in spin structure of eigenstates in the
layers results in a sort of interference in the tunneling
conductance. The dependence of tunneling conductance on voltage
appears to be very sensitive to the parameters of SOI. Thus, we
propose a way to extract the parameters of SOI and, in particular,
the relation between Rashba and Dresselhaus contributions in the
tunneling experiment. We emphasize that unlike many other
spin-related experiments the manifestation of SOI studied in this
paper should be observed without external magnetic field. Our
calculations show that the interference picture may be well resolved
for GaAs samples with the scattering times down to $\sim 10^{-12}$
s, in some special cases the scattering even restores the traces of
SOI otherwise not seen due to destructive interference.
\section*{ACKNOWLEDGEMENTS}
 This work has been supported in part by RFBR, President
of RF support (grant MK-8224.2006.2) and Scientific Programs of RAS.
\appendix*
\section{}
\begin{figure}[h]
 \leavevmode
 \centering\epsfxsize=230pt \epsfbox[89 700 395 780]{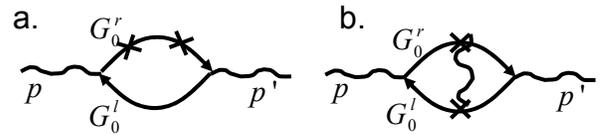}
 \caption{\label{fig:diagrams}Two types of diagrams for calculation of the tunneling current}
\end{figure}

In this section we discuss an approach to the calculation of the
tunneling current based on Green's function formalism. As we shall
see this approach used in\cite{MacDonald,Raikh} gives the same
results as that based on the operator motion equation (\ref{drodt}).
However, one should be accurate with averaging over spatial
distribution of the impurities. The starting point for this
calculation is the tunneling Hamiltonian (\ref{HT0}) in which the
coupling between the two layers sits merely in the tunneling term
$H_T$ (\ref{eqH}). Whenever such Hamiltonian is assumed, its part
connected to the scattering at the impurities is the following
\begin{equation}
\label{scatteringHamiltonian} H_V=H_V^L+H_V^R=
\sum\limits_{k,k',\sigma}  {V^{L}_{kk'}
c^{L+}_{k\sigma}c^L_{k'\sigma }}+\sum\limits_{k,k',\sigma}
{V^R_{kk'} c^{R+}_{k\sigma}c^R_{k'\sigma }},
\end{equation}
where $V^l_{kk'}$ are the matrix elements of the scattering operator
calculated on the electrons eigenfunctions. Note that this
Hamiltonian does not contain interlayer matrix elements
$V^{LR},\;V^{RL}$, which would have appeared after straightforward
secondary quantization of the impurities external field
(\ref{ImpuritiesPotential}). This is reasonable because these
elements are parametrically small compared to intralayer elements
$V^{L},\;V^{R}$ and also to the tunneling term $H_T$. To prove the
first we recall from (\ref{correlators1}) that $V^{ll'}\sim\sqrt{B}$
while the intra-layer matrix elements $V^{l}\sim\sqrt{A}$. Taking
into account (\ref{T2}) we have:
\[
\frac{V^{ll'}}{V^{l}}\sim t\ll1
\]
The interlayer matrix elements are also parametrically small
compared to the tunneling term in the Hamiltonian. Indeed, from
(\ref{tau}),(\ref{T2}) also follows that $V^{ll'}\sim
t\frac{\hbar}{\tau}$ ($l \neq l'$), while as was mentioned earlier
$T\sim t\varepsilon_F$, hence:
\[
\frac{V^{ll'}}{T}\sim \frac{\hbar}{\varepsilon_F\tau}\ll 1
\]
The Hamiltonian (1) can be rewritten as:
\begin{equation}
\label{HPerturbations} H=H_0+H_V+H_T
\end{equation}
The calculation of the current can be carried out by means of
perturbation theory with $H_V$,$H_T$ treated as two perturbations.
At first step only the tunneling term $H_T$ is considered as a
perturbation while the rest of the Hamiltonian describes the system
treated as unperturbed. The tunneling constant $T$ is a small
parameter as soon as our consideration is restricted to the case of
weak tunneling. The d.c. tunneling current can be calculated
adjusting the Kubo formula to the tunneling current\cite{Mahan}. The
Kubo formula treats the perturbation to the leading order so that
the leading order for the current is $T^2$. We obtain the following
formula for the d.c. tunneling current expressed through the
unperturbed Green'a functions:
\begin{equation}
 I = \frac{{eT^2
W}}{{\hbar ^3 }}{\mathop{\rm Im}\nolimits} \left\{ \mathrm{Tr} \int
{ {G_{0V}^R \left( {{p,p},\varepsilon  - eU} \right)G_{0V}^L \left(
{{p,p},\varepsilon } \right)d{\bf{p}}d\varepsilon } }. \right\},
\end{equation} where $p$ is electron lateral momentum, $G_{0V}^R, G_{0V}^L$ are the
 components of the unperturbed Green's functions (which
are matrices in our case). Index $V$ tells that at this stage, while
only $H_T$ is considered as a perturbation, these unperturbed
Green's functions include the scattering by impurities. At the
second step the Green's functions $G_{0V}^l$ are to be expressed in
terms of the known unperturbed Green's functions $G_{0}^l$ of the
non-interacting 2D electron gas.
 The conventional perturbation theory leads to the summation of diagrams of the two
types shown in Fig.\ref{fig:diagrams}
The branches correspond to the unperturbed Green's functions
$G_{0}^l$ while the crosses correspond to the matrix elements
$V_{kk'}^l$. The ladder diagrams of type b. give vertex corrections
discussed in \cite{MacDonald,Raikh}. However, in our opinion, the
ladder diagrams do not give any significant impact to the tunneling
current. It can be easily shown that if the interaction part of the
Hamiltonian is as given by (\ref{scatteringHamiltonian}), the upper
branches of the diagrams in Fig.\ref{fig:diagrams} (corresponding to
$G_0^R$) contain only $V_{kk'}^R$ while the lower branches have only
$V_{kk'}^L$. Hence the diagrams of type a. contain the quadratic
forms of the intralayer matrix elements $A_{kk'}$
(\ref{correlators}) while the diagrams of type b. contain only
interlayer elements $B_{kk'}$. It follows from (\ref{T2}) that after
 averaging over the impurities (with any degree of their spatial correlation) the interlayer
 matrix elements give higher order with regard to the tunneling parameter $T$ and therefore must be omitted
in the leading order calculations.

\end{document}